\documentclass[preprint,review]{elsarticle}

\usepackage{graphicx}
\usepackage{epstopdf}
\usepackage{color}
\usepackage{placeins}
\usepackage{subfig}
\usepackage{amsfonts}
\usepackage{mathrsfs}
\usepackage{graphicx}
\usepackage{multicol}
\usepackage{multirow}
\usepackage{algorithm}
\usepackage{algorithmic}
\usepackage{setspace}
\usepackage{anysize}

\usepackage{amssymb}
\usepackage{amsmath}
\usepackage{amsthm}

\journal{Journal of Computational Physics}

\begin{document}

\begin{frontmatter}



\title{Vector tomography for reconstructing electric fields with non-zero divergence in bounded domains}


\author[label1,label2]{Alexandra Koulouri\corref{cor1}}
\cortext[cor1]{Corresponding author. Tel.: +49 251 8335127 Fax: +49 251 8332729}
\ead{koulouri@uni-muenster.de}
\author[label2]{Mike Brookes}
\author[label3,label4]{Ville Rimpil\"ainen}

\address[label1]{Institute for Computational and Applied Mathematics, University of M\"unster, Einsteinstrasse 62, D-48149 M\"unster, Germany}
\address[label2]{Department of Electrical and Electronic Engineering, Imperial College London, Exhibition Road, London SW7 2BT, United Kingdom}
\address[label3]{Institute for Biomagnetism and Biosignalanalysis, University of M\"unster, Malmedyweg 15, D-48149 M\"unster, Germany}
\address[label4]{Department of Mathematics, University of Auckland, Private bag 92019, Auckland 1142, New Zealand}

\begin{abstract}

In vector tomography (VT), the aim is to reconstruct an unknown multi-dimensional vector field using line integral data. In the case of
a 2-dimensional VT, two types of line integral data are usually required. These data correspond to integration of the parallel and perpendicular
projection of the vector field along the integration lines and are called the longitudinal and transverse measurements, respectively. In most cases,
however, the transverse measurements cannot be physically acquired. Therefore, the VT methods are typically used to reconstruct divergence-free (or
source-free) velocity and flow fields that can be reconstructed solely from the longitudinal measurements. In this paper, we show how vector fields
with non-zero divergence in a bounded domain can also be reconstructed from the longitudinal measurements without the need of explicitly evaluating
the transverse measurements. To the best of our knowledge, VT has not previously been used for this purpose. In particular, we study low-frequency,
time-harmonic electric fields generated by dipole sources in convex bounded domains which arise, for example, in electroencephalography (EEG) source
imaging. We explain in detail the theoretical background, the derivation of the electric field inverse problem and the numerical approximation of the
line integrals. We show that fields with non-zero divergence can be reconstructed from the longitudinal measurements with the help of two sparsity
constraints that are constructed from the transverse measurements and the vector Laplace operator. As a comparison to EEG source imaging, we note
that VT does not require mathematical modelling of the sources. By numerical simulations, we show that the pattern of the electric field can be
correctly estimated using VT and the location of the source activity can be determined accurately from the reconstructed magnitudes of the field.

\end{abstract}


\end{frontmatter}



\section{Introduction}

Vector fields such as gravitational and electromagnetic fields are fundamental objects of study in physics. Vector tomography (VT) is a framework
that can be used to reconstruct such unknown vector fields using line integral measurements \cite{Schuster2008,Sparr1998}. The longitudinal line
measurements are obtained by projecting the studied field on lines that trace the domain and then integrating the projected field along the lines.
The transverse line measurements are acquired similarly but now the field components that are perpendicular to these lines are integrated. VT methods
are attractive because they can be used with non-invasive  measurement techniques (e.g. ultra-sound, \cite{Jovanovia2009,Johnson1997}) that can give
a larger amount of data \cite{Giannakidis2010,Koulouri2012} compared to the one-sensor one-measurement set-up \cite{Nunez2006}.

VT studies have been carried out both in theoretical level and applications, concentrating mainly on the reconstruction of smooth vector fields
\cite{Sparr1998}. Theoretical analysis for the reconstruction of smooth velocity fields have been presented in \cite{Norton1988,
Norton1992,Schwarz1995,Johnson1975,Braun1991,Lovstakken2006,Rouseff1991,Wernsdorfer1993,Kramar2005,Javanovic2008} using such methods as the inverse
Radon transform \cite{Helgason1999}, the inverse Fourier transform with central slice theorem \cite{Natterer2001,Deans2007} and back projection
(parallel beam tomography) \cite{Lade2005,Schuster2008a}. The VT framework has been used for the reconstruction of particle distributions
\cite{Balandin1989}, ion fields in plasma \cite{Efremov1995,Howard1996,bal05,Balandin2012}, velocity fields in blood veins
\cite{Johnson1975,Johnson1979,Sparr1998}, magnetic field of the corona of the sun \cite{Kramar2005}, Kerr effect in optical polarization tomography
\cite{Hertz1986} and micro-structures in oceanographic tomography \cite{Rouseff1991}. Both linear and non-linear iterative algorithms have been
proposed for vector functions with appropriate smoothness
\cite{Wernsdorfer1993,Natterer2001,Javanovic2008,Petrou2010,Giannakidis2010,Koulouri2012,Sparr1998}.

\subsection{Unbounded domain}

The theoretical basis for reconstructing smooth vector fields that decay sufficiently rapidly to zero in the spatial domain was introduced in
\cite{Norton1988}. Based on Helmholtz's theorem \cite{Arfken2005}, vector fields can be decomposed as a sum of irrotational (curl-free) and
solenoidal (divergence-free or source-free) components and it was first shown that, for a 2-dimensional field, the solenoidal component can be imaged
with the help of longitudinal measurements \cite{Norton1988}. It was subsequently shown that the transverse measurements were required in order to
recover the remainder of the field \cite{Braun1991}.

The problem was extended to three dimensions in \cite{Prince1994} using the formalism of the 3-dimensional (3D) vector Radon transform. First, a
generalization of the integral measurement was introduced. It was called the probe transform (or general product measurement) and it was formulated
as an inner product between the Radon transform of a field and a unit-vector in a specific direction. It was also shown that three types of
measurements were required for the recovery of a 3D field. In \cite{Sparr1998}, the analysis was generalized to multidimensional cases.

However, in most practical situations,
 it is difficult or even impossible to perform the transverse measurements
 (i.e. the probe transform in the transverse direction). For example in Doppler techniques \cite{Desbat1995} or in geophysics \cite{Norton1992}, this type of measurement is not physically
realizable. In fact, the transverse measurements can be obtained only in very specific set-ups \cite{Braun1991,Jovanovia2009}.

\subsection{Bounded domain}

In practical applications, vector fields are defined in bounded domains where the field is not identically zero at the boundary. In fact, it is often
the boundary that partially defines the field itself: for example, the homogeneous Neumann condition implies that the field can have only tangential
component on the boundary. The VT framework was extended to non-homogenous boundary conditions in \cite{Braun1991,Osman1997}. The field decomposition
included an additional harmonic field component that satisfied the boundary conditions \cite{Braun1991}. In 2D circular domains, it was found that
the harmonic component is imaged equally in both the longitudinal and transverse measurements but it had half of its magnitude \cite{Braun1991}. In
\cite{Osman1997}, the results were generalized in 3D arbitrary shaped domains. In particular, it was shown that the longitudinal measurement can be
used to image both the homogeneous solenoidal component and the part of the harmonic term that arises from the field component that is tangential to
the boundary. Additionally, transverse measurement reconstructs the irrotational component and the harmonic part that arises from the field component
that is normal to the boundary.

\subsection{Electric field with non-zero divergence}

There are theoretical studies in which arbitrary vector fields have been investigated \cite{Schuster2008}. However, to the best of our knowledge,
there are no previous studies in which numerical reconstructions of non-zero divergence vector fields in a bounded domain have been carried out using
only the longitudinal line measurements. This kind of vector fields are common in physics and can be, for example, gravitational or electromagnetic
fields that are generated by unknown sources (and/or sinks) that are located inside the domain of interest. In this paper, the aim is to use VT to
reconstruct such non-zero divergence vector fields. In particular, we employ VT for the reconstruction of low-frequency, time-harmonic electric
fields in a convex bounded domain that includes a dipole source. Strategies to estimate such electric activity are of great interest especially in
medical imaging modalities such as electroencephalography (EEG) in which the imaging problem is traditionally parametrisized using source spaces
\cite{Hamalainen1993,Wolters2004}. The proposed VT modelling assumes the same physical conditions as the dipole source imaging problem i.e. the
underlying electric field is irrotational. The existence of a dipole inside the domain implies that the field has a singularity. Previously, it has
been shown that VT can be used also in such cases \cite{Derevtsov2008,Derevtsov2011}.

The use of VT rather than traditional inverse source approaches \cite{Grech2008,Nunez2006}
offers two advantages. First, the continuous VT problem for the recovery of the electric
field using a set of line integrals is only a
moderately ill-posed problem \cite{Natterer2001} whereas the inverse
source problem is a severely ill-posed problem that cannot be solved
from boundary measurements without a priori knowledge
\cite{Antoniadis2006}. In practice, however, prior information is also required by the VT formulation (e.g. introduced as a penalty term) in order to obtain a stable
solution since only a finite number of measurements is available for the reconstruction (incomplete data problem \cite{Natterer2001}). Second, the VT
approach does not require an explicit mathematical model of the underlying sources. For example, in EEG source imaging there is an extensive
literature on different mathematical models of neural sources \cite{Johnson1993,Johnson1997,Buchner1997,Huiskamp1999,Schimpf2002,Nunez2006,bau15}.

In the proposed VT approach, we use a set of line integrals that trace a conductive 2-dimensional domain and result in a linear system of equations.
We show that the longitudinal measurements are determined by the electric potentials on the domain boundary and by employing the vectorial Radon
properties and the homogeneous Neumann boundary condition that the transverse measurements give information on the underlying current sources. We describe in detail the theoretical background, the
numerical approximation of the line integrals and finally present the discretized electric field inverse problem which is solved with the help of the
$L_1$-norms of the transverse measurements and the discrete vector Laplace operator.  The resulting non-linear minimization problem is solved using
convex optimization. Finally, we show by numerical simulations that electric fields with non-zero divergence can be reconstructed in a bounded domain.

\section{Mathematical preliminaries}

In this section, we explain the notations and define the function spaces and the different Radon measurements. More general information on the Radon
transform can be found from \cite{Helgason1999,Deans2007}.

\subsection{Distributions}

The theory of distributions (a.k.a. generalized functions) provides a powerful framework to describe the potentials and fields of the electromagnetic
theory \cite{Skinner1989}. It allows one to calculate such physical quantities as point dipoles and electric fields with singularities and/or
discontinuities which cannot be estimated using the classical calculus \cite{Skinner1989}. Therefore, in the following analysis we consider that the
studied electromagnetic quantities belong to the space of distributions denoted by $\mathcal{E}'(\mathbb{R}^d;\mathbb{R}^2)$ in the unbounded domain
$\mathbb{R}^2$ and $\mathcal{E}'(\mathbb{R}^d;\Omega)$ in $\Omega$ which is convex, open and bounded with smooth boundary $\partial\Omega$
\cite{Sharafutdinov1994}. Here, the index $d$ denotes the dimension of the function i.e. $d=1$ for scalars and $d=2$ for vector valued functions
which are denoted by $f$ and $\mathrm{f}$, respectively. Moreover, the values (or measures) of $f$ are given by the scalar product $\langle f,
\varphi \rangle$ where $\varphi\in \mathcal{C}_0^\infty$ is a set of smooth, compactly supported (usually localized) test functions defined based on
the properties of the electromagnetic problem \cite{Skinner1989}. Accordingly, in the current problem the Radon transforms will be interpreted in the
sense of distributions \cite{Deans2007,Osman1997}.

\subsection{Radon transform of a scalar function}
\label{sdefs}

\begin{figure}[!htb]\centering
\includegraphics[width=9cm]{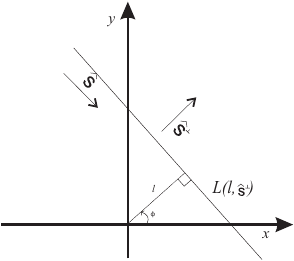}
\caption{The integration path $L$ is defined with the help of the signed distance from the origin $l$ and the unit normal vector $\hat{s}^{\perp}$.
In addition, $\hat{s}$ denotes the unit vector along $L$.} \label{fig:prel}
\end{figure}

We denote by $\mathrm{x} \in \mathbb{R}^2$ a point, $f:\mathbb{R}^2\rightarrow \mathbb{R}$ a scalar function and
$L(l,\mathrm{\hat{s}}^{\perp}):=\{\mathrm{x}=(x,y) \in \mathbb{R}^2:\mathrm{x}\cdot\mathrm{\hat{s}}^{\perp}=l\}$ a line where ${l} \in \mathbb{R}$ is
the signed distance of the line from the origin and $\mathrm{\hat{s}}^\perp = (\cos \phi,\sin\phi)$ is the unit normal vector of $L$ (see, Figure
\ref{fig:prel} for details). The angle $\phi \in [0,\,2\pi)$ is measured counter-clockwise from the positive x-axis. Similarly, we define
$\mathrm{\hat{s}} = (-\sin \phi,\cos \phi)$ that is the unit vector parallel to the line $L$.

The mapping defined by the line integral of $f(\mathrm{x})$ along a line $L$ is the two-dimensional Radon transform of $f$ and is given by
\begin{equation}
\tilde{f} =\mathcal{R}\{f\}(l,\hat{s}^{\perp}) = \int_{\mathrm{x} \in L} f(\mathrm{x})~d\ell(\mathrm{x}),
\end{equation}
where $d\ell(\mathrm{x})$ is an increment of length along $L$ \cite{Natterer2001} and $\mathcal{R}: \mathcal{E}'(\mathbb{R};\mathbb{R}^2) \rightarrow
\mathcal{D}'(\mathbb{R} \times [0,\,2\pi))$ where $\mathcal{D}'$ denotes the space of symmetric distributions \cite{Sharafutdinov1994,Deans2007}. If
the position vector on $L$ is described by $\mathrm{x}=l\hat{\mathrm{s}}^\perp+t\hat{\mathrm{s}}$ where $t\in\mathbb{R}$, then the line integral can
be written as
$$\tilde{f}=\int_{\mathbb{R}}f(l\hat{\mathrm{s}}^\perp+t\hat{\mathrm{s}})~dt.$$
This equation implies that the integration is always performed in the direction of $\hat{\mathrm{s}}$. Using the Dirac-delta function
\cite{Deans2007}, the line integral can be expressed as the surface integral
\begin{equation}
\tilde{f}=\mathcal{R}\{{f}\}(l,\mathrm{\hat{s}}^\perp)=
\int_{\mathbb{R}^2}{f}(\mathrm{x})\delta(l-\mathrm{x}\cdot\mathrm{\mathrm{\hat{s}}}^{\perp})~d\mathrm{x}.
\end{equation}

The corresponding scalar inverse Radon transform is
\begin{equation}\label{eq:InverseRadon}
f=\mathcal{R}^{-1}\{\tilde{f}\} = \frac{1}{4\pi}\mathcal{R}^{\#}\mathcal{H}\frac{\partial}{\partial l} \tilde{f},
\end{equation}
where $\mathcal{R}^{\#}$ is the adjoint operator of $\mathcal{R}$, $\mathcal{H}\frac{\partial}{\partial l}$ is the filtration part of the inverse
transform and $\mathcal{H}$ denotes the Hilbert transform \cite{Deans2007}.

Additionally, we will find useful the Radon transform of a directional derivative \cite{Deans2007} along a unit vector $\mathrm{\hat{\omega}}$
\begin{equation}\label{eq:Radon_derivative}
\mathcal{R}\{\mathrm{\hat{\omega}}\cdot\nabla f\}= \mathrm{\hat{\omega}}\cdot\mathrm{\hat{s}}^{\perp}\frac{\partial}{\partial l} \mathcal{R}\{{f}\},
\end{equation}
which is valid for $f, \nabla f \in \mathcal{E}'$ and $\mathcal{R}\{\mathrm{\hat{\omega}}\cdot\nabla f\} \in \mathcal{D'}$ because $\langle
\mathcal{R}\{\mathrm{\hat{\omega}}\cdot\nabla f\}, \varphi \rangle = \langle
\mathrm{\hat{\omega}}\cdot\mathrm{\hat{s}}^{\perp}\frac{\partial}{\partial l} \mathcal{R}\{{f}\}, \varphi \rangle \, \, \forall \varphi \in
\mathcal{C}_0^\infty$ \cite{Deans2007}.

\subsection{$\Omega$-Radon transform of a scalar function}
 In real applications usually we consider that $f:\Omega\rightarrow \mathbb{R}$
 where $\Omega$ is a simply connected domain, open and bounded with smooth boundary $\partial\Omega$. The  $\Omega$-Radon transform is
\begin{equation}
\tilde{{f}}_{\Omega}= \mathcal{R}_{\Omega}\{{f}\}(l,\mathrm{\hat{s}}^{\perp})  =\int_{\Omega}
{f}(x)~\delta(l-\mathrm{x}\cdot\mathrm{\hat{s}}^{\perp})~d\mathrm{x}.
\end{equation}
If $f^c \in \mathcal{E}'$, defined on $\mathbb{R}^2$, is such that ${f}^c={f}$ on $\Omega$, then $\Omega$-Radon transform $\mathcal{R}_\Omega$ is the
restriction of the Radon transform to functions on $\mathbb{R}^2$ that identically vanish outside of $\Omega$ expressed as
\begin{equation}
\label{extendingRadon}
\tilde{{f}}_{\Omega}= \int_{\mathbb{R}^2} {f}^c(\mathrm{x})~v(h(\mathrm{x}))~\delta(l-\mathrm{x}\cdot\mathrm{\hat{s}}^{\perp})~d\mathrm{x}= \mathcal{R}\{f^c
v_{\Omega}\}(l,\mathrm{\hat{s}}^{\perp}).
\end{equation}
For simplicity, we write $v_\Omega=v(h(\mathrm{x}))$. The indicator(Heaviside) function $v:\mathbb{R}\rightarrow\mathbb{R}$ is defined as
\begin{equation}
    v(s)=
\begin{cases}
    1,& s> 0\\
    0,& s\leq 0
\end{cases}
\end{equation}
\noindent where $h(\mathrm{x})$ is a signed distance function satisfying
\begin{equation}
    h(\mathrm{x})=
\begin{cases}
    d(\mathrm{x},\partial\Omega),& \mathrm{x}\in(\Omega\bigcup\partial\Omega)\\
    -d(\mathrm{x},\partial\Omega),& \mathrm{x}\in\mathbb{R}^2/(\Omega\bigcup\partial\Omega)
\end{cases}
\end{equation}
where $d(\mathrm{x},\partial\Omega):=\inf_{y\in\partial\Omega}d(x,y)$ is the shortest distance of the point $\mathrm{x}$ to the boundary $\partial
\Omega$. For the signed distance function, we have that $\nabla h(\mathrm{x})=-\hat{n}(\mathrm{x})$ on a (piecewise) smooth boundary
$\partial\Omega$, where $\hat{n}$ is the outward unit normal vector \cite{Osher2003}.
\subsection{$\Omega$-Radon transform of a vector function}
In the current analysis, we operate in a bounded convex domain $\Omega\subset \mathbb{R}^2$ on a vector function $\mathrm{f}=(f_x,f_y): \Omega
\rightarrow \mathbb{R}^2$. The vectorial Radon transform of vector $\mathrm{f}$ is the Radon transform of its elements \cite{Osman1997}, i.e.
\begin{equation}
\tilde{\mathrm{f}}_\Omega = (\tilde{f}_{\Omega x},\tilde{f}_{\Omega y})=\mathcal{R}_\Omega\{\mathrm{f}\}(l,\mathrm{\hat{s}}^\perp).
\end{equation}
As we will see in the next section, the inner product of the vectorial Radon transform with a unit vector yields to a scalar quantity which can be
measured in some applications.

\subsection{Line integral data}

In 2-dimensional VT, two different types of line integral measurements are used to reconstruct vector fields. The first is the line integral
\begin{equation}\label{eq:lineIntegralPreviousWork}
I_L^{\parallel}(l,\hat{\mathrm{s}}^{\perp})=\int_{L}
\hat{\mathrm{s}}\cdot\mathrm{f}(\mathrm{x})~d\ell(\mathrm{x})=\hat{\mathrm{s}}\cdot\mathcal{R}_{\Omega}\{\mathrm{f}\}(l,\mathrm{\hat{s}}^\perp),
\end{equation}
\textcolor[rgb]{0,0,0}{which is the product of the vectorial Radon with the unit vector $\hat{s}$} and called the longitudinal measurement \cite{Schuster2008}. The second line integral is called the transverse measurement \cite{Schuster2008} and it is defined as
\begin{equation}\label{eq:lineIntegralPreviousWork_transverse}
I_L^{\perp}(l,\mathrm{\hat{s}}^{\perp})=
\int_{L}\hat{\mathrm{s}}^\perp\cdot\mathrm{f}(\mathrm{x})~d\ell(\mathrm{x})=\hat{\mathrm{s}}^\perp\cdot\mathcal{R}_{\Omega}\{\mathrm{f}\}(l,\mathrm{\hat{s}}^\perp).
\end{equation}
The unit vectors $\hat{\mathrm{s}}=(s_x,s_y)$ and $\hat{\mathrm{s}}^{\perp}=(-s_y,s_x)$ are defined as in Section \ref{sdefs} and Figure
\ref{fig:prel}.

\section{Theory}
\label{th1}

\subsection{Electric field with a current source in a bounded domain}

Let us consider a bounded convex domain $\Omega \subset \mathbb{R}^2$ with electrical conductivity $\sigma(\mathrm{x})$, $\mathrm{x}\in\Omega$ and an
electric source with (primary) current density $\mathrm{j}^s:\Omega\rightarrow \mathbb{R}^2$. The electric source induces an electric field
$\mathrm{e}:\Omega\rightarrow\mathbb{R}^2$. The total current density in the medium can be presented as a sum of the primary and secondary current
\cite{Kip69}, i.e.
\begin{equation}
\mathrm{j}(\mathrm{x})  =\mathrm{j}^s(\mathrm{x})+
\sigma(\mathrm{x})\mathrm{e}(\mathrm{x}).\label{eq:Appendix_Current_LI}
\end{equation}

For current signals with low frequencies, the
quasi-static Maxwell equations can be used
\begin{subequations}
\label{eq:MaxwellEquations_harmonic}
\begin{align}\label{eq:Appendix_Maxwell_1_h}
\nabla\times \mathrm{e}(\mathrm{x}) &= 0\\
\label{tokaMax}
\nabla \times \mathrm{h}(\mathrm{x}) &=\mathrm{j}(\mathrm{x}),
\end{align}
\end{subequations}
where $\mathrm{h}(\mathrm{x})$ is the magnetic field intensity.
The divergence of equation (\ref{tokaMax}) gives
\begin{subequations}
\begin{align}
\label{efield}
\nabla \cdot \nabla \times \mathrm{h}(\mathrm{x}) &=\nabla \cdot \mathrm{j}(\mathrm{x})\\
0 &= \nabla \cdot (\mathrm{j}^s(\mathrm{x})+
\sigma(\mathrm{x})\mathrm{e}(\mathrm{x})) \\
\nabla\cdot\sigma\mathrm{e}(\mathrm{x})&
=-\nabla\cdot\mathrm{j}^s(\mathrm{x}),\label{eq:FieldSourcePoisson}
\end{align}
\end{subequations}
which relates the electric field to the current source.

Because the electric field is irrotational, Equation (\ref{eq:Appendix_Maxwell_1_h}), the field can be expressed as
\begin{equation}
\mathrm{e(x)} = - \nabla u
\end{equation}
\noindent where $u$ is a scalar potential \cite{Hamalainen1993,Wolters2004}. In this paper, we consider that $u$ is
uniquely defined as the solution of the Poisson equation with the following boundary conditions
\begin{subequations}
\begin{align}
\label{p1}
\nabla \cdot \sigma \nabla u &= \nabla\cdot\mathrm{j}^s(\mathrm{x}) \\
\label{p3} \frac{\partial u}{\partial \hat{n}} &= 0, \, \mbox{on  } \partial \Omega,\\
\label{p2} \int_{\partial \Omega} u ~dS&= 0,
\end{align}
\end{subequations}
\noindent where $\partial \Omega$ is the boundary and $\hat{n}$ is the outward unit normal vector. The homogeneous Neumann condition (\ref{p3})
implies that the electric field is tangential at the boundary,  $\hat{n} \cdot \mathrm{e} |_{\partial \Omega}=0$, and the condition (\ref{p2})
ensures that the solution is unique \cite{Malmivuo1995,Arfken2005}.

\subsection{Line integrals through direct substitution}
\label{dirsub}

In this vector tomography framework, we consider two types of line integral measurements of the electric field. If we directly evaluate the integrals (\ref{eq:lineIntegralPreviousWork}) and (\ref{eq:lineIntegralPreviousWork_transverse}) by substituting the electric field with the negative gradient of the scalar potential, first, the longitudinal integral measurements are
\begin{equation}
\label{eq:model equation_long} I_{L(\mathrm{x}_a,\mathrm{x}_b)}^{\parallel}=\int_{L(\mathrm{x}_a,\mathrm{x}_b)} \mathrm{e}(\mathrm{x})\cdot
\hat{\mathrm{s}}~d\ell(\mathrm{x}) = \int_{L(\mathrm{x}_a,\mathrm{x}_b)} -\nabla u \cdot
\hat{\mathrm{s}}~d\ell(\mathrm{x}) =u(\mathrm{x}_a)-u(\mathrm{x}_b),
\end{equation}
\noindent where $u(\mathrm{x}_a)$ and $u(\mathrm{x}_b)$ are the electric potential values at the intersections of the line $L$ and the boundary
$\partial\Omega$ and $\hat{\mathrm{s}}=(s_x,s_y)$ is the unit vector as defined in Section \ref{sdefs} and Figure \ref{fig:prel}. Second, the
transverse line integral measurements are
\begin{subequations}
\begin{align} \label{eq:model equation_trav}
I_{L(\mathrm{x}_a,\mathrm{x}_b)}^{\perp}&=\int_{L(\mathrm{x}_a,\mathrm{x}_b)} \mathrm{e} (\mathrm{x})\cdot\hat{{\mathrm{s}}}^{\perp}~d\ell(\mathrm{x})\\
&=\int_{L(\mathrm{x}_a,\mathrm{x}_b)} -\nabla u \cdot \hat{\mathrm{s}}^{\perp}~d\ell(\mathrm{x})\\
&=\int_{L(\mathrm{x}_a,\mathrm{x}_b)} -\left( \frac{\partial u}{\partial x},\frac{\partial u}{\partial y} \right) \cdot (-s_y,s_x)~d\ell(\mathrm{x})\\
&=\int_{L(\mathrm{x}_a,\mathrm{x}_b)} \left( \frac{\partial u}{\partial x} s_y-\frac{\partial u}{\partial y}s_x \right) ~d\ell(\mathrm{x}).
\end{align}
\end{subequations}

As it can be seen, the longitudinal integral measurements are directly determined by the boundary potentials; however, the transverse integral
measurements in practice cannot be measured directly (or only under special circumstances \cite{Braun1991,Jovanovia2009}). This turns out to be a
problem because the full recovery of the electric field requires both types of integral measurements. In the Appendix,
it is shown that only the harmonic component of the electric field can be reconstructed from the longitudinal line integrals, whereas the remaining
irrotational part requires the transverse measurements. The transverse integral formulations nevertheless turn out to be useful since, as will be
seen in the following section, they give information about the underlying current sources that generate the field.

\subsection{Transverse line integral through $\Omega$-Radon transform}\label{sec:TransverseThroughRadon}

In this section, we show that the transverse line integral measurement is  related directly to the underlying current source when the homogenous
Neumann condition holds. We start by taking the $\Omega$-Radon transform of both sides of Equation (\ref{eq:FieldSourcePoisson}). For simplicity, we
consider that the electric conductivity is constant.
\begin{subequations}
\begin{align}
\nabla \cdot \sigma \mathrm{e(x)}&=-\nabla \cdot \mathrm{j^s(x)} \\
\sigma \mathcal{R}_\Omega \{\nabla \cdot \mathrm{e(x)} \}&=- \mathcal{R}_\Omega \{\nabla \cdot \mathrm{j^s(x)} \}.\label{eq:radondivergences}
\end{align}
\end{subequations}
Similarly as in \cite{Osman1997}, we define the extension of $\mathrm{e}^c$ in $\mathbb{R}^2$ such that $\mathrm{e}^c=\mathrm{e}$ in $\Omega$ in
order to utilize the Radon property (\ref{extendingRadon}) as follows
\begin{equation}
\label{eq:radondiv}
 \mathcal{R}_\Omega \{\nabla \cdot \mathrm{e} \} = \int_{\mathbb{R}^2} (\nabla \cdot \mathrm{e}^c) ~v_{\Omega} \delta(\ell-\mathrm{x}\cdot\hat{\mathrm{s}}^{\perp}) d\mathrm{x}.
\end{equation}
Also, the divergence $\nabla\cdot \mathrm{e}^cv_\Omega$ equals to
\begin{equation}
\label{eq:div1}
 \nabla\cdot(\mathrm{e}^cv_{\Omega})=(\nabla\cdot\mathrm{e}^c)~v_{\Omega}+(\nabla h)\cdot\mathrm{e}^c~\delta_{\partial\Omega},
\end{equation}
where $\delta_{\partial\Omega}=\delta(h(\mathrm{x}))$. From the gradient of the signed distance function and the boundary condition (\ref{p2}), we
get $\nabla h\cdot e|_{\partial \Omega} = -\hat{n}\cdot\mathrm{e}|_{\partial\Omega}=\frac{\partial u}{\partial \hat{n}} |_{\partial \Omega}=0$. Now,
we can re-write Equation (\ref{eq:div1})
\begin{equation}
\nabla\cdot(\mathrm{e}^cv_{\Omega})=(\nabla\cdot\mathrm{e}^c)~v_{\Omega}.
\end{equation}
So, Equation (\ref{eq:radondiv}) becomes
\begin{equation}
\mathcal{R}_\Omega \{\nabla \cdot \mathrm{e} \}=\mathcal{R}\{\nabla \cdot( \mathrm{e}^cv_\Omega) \}
\end{equation}
Using  property (\ref{eq:Radon_derivative}), i.e. $\mathcal{R}\{\nabla \cdot( \mathrm{e}^cv_\Omega)
\}=\hat{\mathrm{s}}^\perp\cdot\frac{\partial}{\partial l}\mathcal{R} \{\mathrm{e}^cv_\Omega \}$, we finally have that
\begin{equation}
 \mathcal{R}_\Omega \{\nabla \cdot \mathrm{e} \} =\hat{\mathrm{s}}^\perp\cdot\frac{\partial}{\partial l}\mathcal{R}_ \Omega\{\mathrm{e} \}.
\end{equation}
Similarly, under the assumption that the (extended) source function is zero outside the domain $\Omega$, we obtain
\begin{equation}
 \mathcal{R}_\Omega \{\nabla \cdot \mathrm{j}^s \} =\hat{\mathrm{s}}^\perp\cdot\frac{\partial}{\partial l}\mathcal{R}_ \Omega\{\mathrm{j}^s \}.
\end{equation}
Hence, Equation (\ref{eq:radondivergences}) is re-written as
\begin{equation}
\hat{\mathrm{s}}^{\perp} \cdot \frac{\partial}{\partial l} \mathcal{R}_\Omega \{\mathrm{e} \}=- \hat{\mathrm{s}}^{\perp} \cdot \frac{1}{\sigma}\frac{\partial}{\partial l} \mathcal{R}_\Omega \{\mathrm{j^s} \}.
\end{equation}
\noindent
 Now, the inverse Radon transform (\ref{eq:InverseRadon}) gives us
\begin{subequations}
\begin{align}
 \mathcal{R}^{-1} \{\mathrm{\hat{s}}^\perp\cdot\mathrm{\tilde{e}_{\Omega}}\}&=
 \frac{1}{4 \pi}
 \mathcal{R}^{\#}\mathcal{H}\frac{\partial}{\partial l}[\mathrm{\hat{s}}^\perp\cdot\mathcal{R}_\Omega\{\mathrm{e}\}] \\
 &= -\frac{1}{4\pi \sigma}
 \mathcal{R}^{\#}\mathcal{H}\frac{\partial}{\partial
l}[\mathrm{\hat{s}}^\perp\cdot\mathcal{R}_\Omega\{\mathrm{j}^s\}]  \\ &=-\frac{1}{\sigma} \mathcal{R}^{-1}\{\mathrm{\hat{s}}^\perp\cdot
\tilde{\mathrm{j}}^s\}.
\label{eq:InverseTranverseInt}
\end{align}
\end{subequations}

\noindent Therefore, we have that the transverse measurements are
\begin{equation}\label{eq:TranverseIntSource}
I_L^{\perp}=\mathrm{\hat{s}}^\perp\cdot\mathcal{R}_{\Omega}\{\mathrm{e}\}
=-\mathrm{\hat{s}}^\perp \cdot \frac{1}{\sigma}
\mathcal{R}_{\Omega}\{\mathrm{j}^s\}.
\end{equation}
In other words, we have shown that the transverse integral
measurements give us information about the source activity inside
the bounded domain.

\subsection{Dipole sources and transverse measurements}
\label{fs}

In this paper, we consider focal source activity $\mathrm{j}^s(\mathrm{x})\in\mathcal{E}'(\mathbb{R}^2;\Omega)$ and dipoles in particular which can be described with the help of Dirac delta functions as
\begin{equation}
\mathrm{j}^s(\mathrm{x}) = \sum_{i=1}^{N_s} \mathrm{q}_i\delta(\mathrm{x}-\mathrm{x}_i),
\end{equation}
where $\mathrm{q}_i$ is the dipole moment, $\mathrm{x}_i$ the dipole source location and $N_s$ the total number of dipole sources \cite{Wolters2003}.

Based on the theory of integral geometry for distributions \cite{Sharafutdinov1994}, the values of the transverse integral $I^{\perp} = -
\frac{1}{\sigma}\mathrm{\hat{s}}^\perp \cdot \mathcal{R}_{\Omega}\{\mathrm{j}^s\}$, when $\mathrm{j}^s(\mathrm{x})$ is given as above, can be
estimated using the following scalar product $\langle\mathrm{\hat{s}}^\perp \mathcal{R}_{\Omega}\{\mathrm{j}^s\}, \varphi \rangle$
$\forall\;\varphi(\mathrm{\hat{s}}^{\perp},l)\in C_c^2( (0,2\pi]\times U)$, $\mathrm{\hat{s}}^{\perp} = (\cos \phi,\sin\phi)$ with $\phi\in(0,2\pi]$
and $l\in U:=\{l =\mathrm{x}\cdot \mathrm{\hat{s}}^{\perp}\; \forall\;\mathrm{x}\in \Omega,\; \phi\in(0,2\pi] \}$ (See Figure 1). This scalar product
gives us
\begin{eqnarray*}
\label{eq:DipoleSourceAndRadonIntegral}
\langle\mathrm{\hat{s}}^\perp\cdot \mathcal{R}_{\Omega}\{\mathrm{j}^s\}, \varphi \rangle
\equiv&\int_U\int_{\phi}\int_{\Omega}\mathrm{\hat{s}}^\perp\cdot\mathrm{j}^s(\mathrm{x})~\delta(l-\mathrm{x}\cdot\mathrm{\hat{s}}^{\perp})
\varphi(\mathrm{\hat{s}}^\perp,l)~d\mathrm{x}\;d\phi\; dl\\
=& \int_{\phi}\int_{\Omega}\mathrm{\hat{s}}^\perp\cdot\mathrm{j}^s(\mathrm{x})
\varphi(\mathrm{\hat{s}}^\perp,\mathrm{x}\cdot\mathrm{\hat{s}}^{\perp})~d\mathrm{x}~d\phi\\
=&\int_{\phi}\int_{\Omega}\mathrm{\hat{s}}^\perp\cdot\left(
\sum_{i=1}^{N_s} \mathrm{q}_i\delta(\mathrm{x}-\mathrm{x}_i)\right)
\varphi(\mathrm{\hat{s}}^\perp,\mathrm{x}\cdot\mathrm{\hat{s}}^{\perp})~d\mathrm{x}\;d\phi\\
=& \int_{\phi}\sum_{i=1}^{N_s} \mathrm{\hat{s}}^\perp\cdot \mathrm{q}_i \varphi(\mathrm{\hat{s}}^\perp,\mathrm{x}_i\cdot\mathrm{\hat{s}}^{\perp})~d\phi\\
=&\sum_{i=1}^{N_s}\int_{k=\mathrm{x}_i\cdot\mathrm{\hat{s}}^{\perp}}
\int_{\phi} \mathrm{\hat{s}}^\perp\cdot \mathrm{q}_i \varphi(\mathrm{\hat{s}}^\perp,k)~d\phi\; dk\\
=&\int \int_{\phi}\sum_{i=1}^{N_s} \mathrm{\hat{s}}^\perp\cdot \mathrm{q}_i\delta(k-\mathrm{x}_i\cdot\mathrm{\hat{s}}^{\perp})
  \varphi(\mathrm{\hat{s}}^\perp,k)~d\phi\; dk\\
 \equiv&\langle\sum_{i=1}^{N_s}\mathrm{\hat{s}}^\perp\cdot \mathrm{q}_i\delta(k-\mathrm{x}_i\cdot\mathrm{\hat{s}}^{\perp}),\varphi\rangle.
\end{eqnarray*}
Therefore, we can write for the transverse measurement that
\begin{equation}\label{eq:IL} I^{\perp} =  -
\frac{1}{\sigma}\mathrm{\hat{s}}^\perp \cdot \mathcal{R}_{\Omega}\{\mathrm{j}^s\}= -\frac{1}\sigma\sum_{i=1}^{N_s}\mathrm{\hat{s}}^\perp\cdot
\mathrm{q}_i\delta(k-\mathrm{x}_i\cdot\mathrm{\hat{s}}^{\perp})\end{equation}

\begin{figure}[!htb]\centering
\includegraphics[width=9cm]{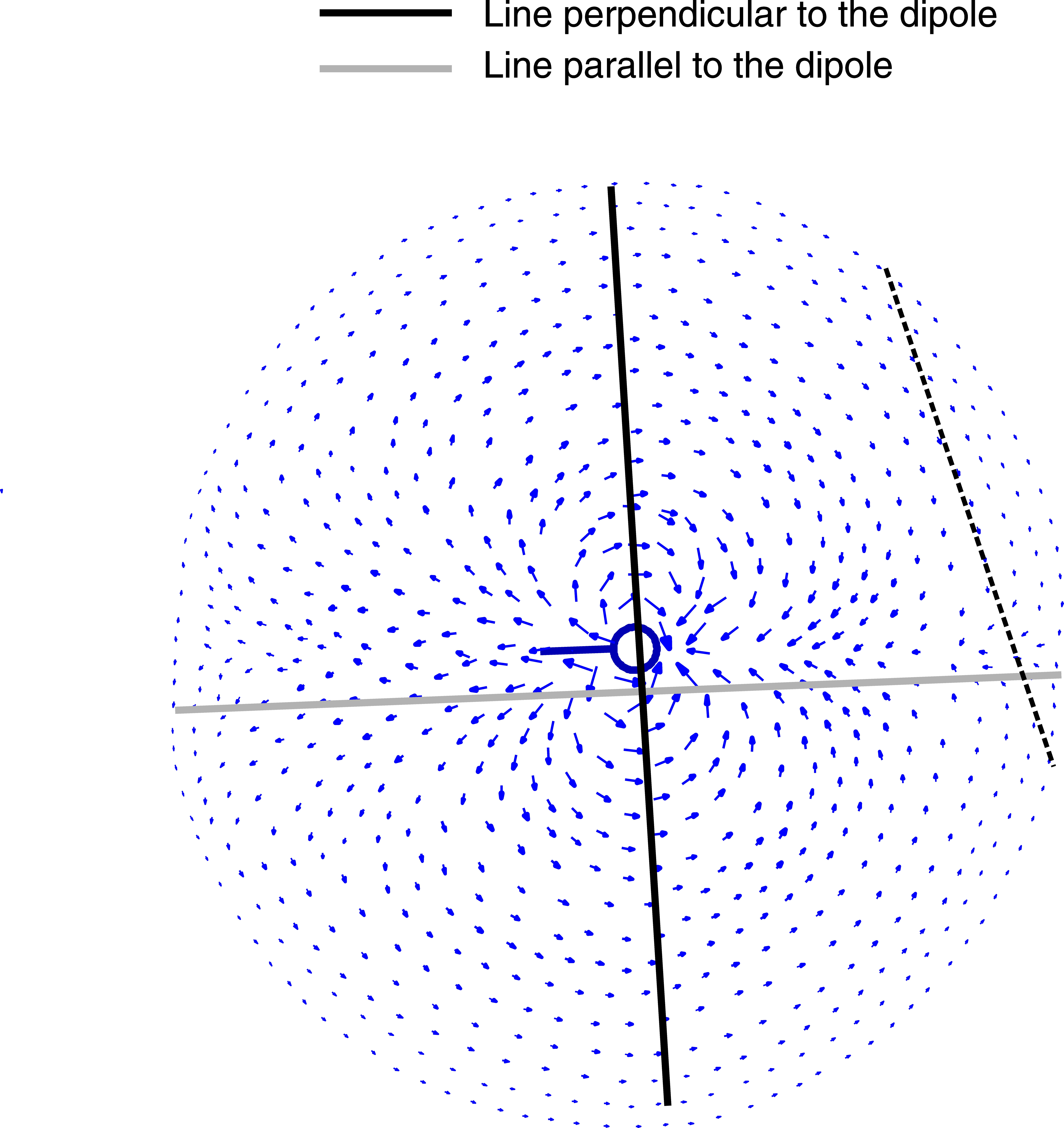}
\caption{Electric field (arrows) that is generated by a dipole source (circle).Of the three shown lines of integration, the gray one is parallel and
the black one perpendicular to the orientation of the dipole. The transverse line integral along the gray line is zero due to the dot-product in
Equation (\ref{eq:IL}), however the transverse integral along the black line is non-zero. The transverse line integral along the dashed line is zero
because the line does not pass through the source, and also the longitudinal line integral is close to zero because the line is far from the dipole
source.} \label{fig:LineOrthogonalToVectorMoment}
\end{figure}

This (\ref{eq:IL}) implies that the transverse measurement is non-zero only when the line of integration passes through the source location and
the line is not parallel to the dipole moment, such as  the black line in Figure~\ref{fig:LineOrthogonalToVectorMoment}. In VT, we have a set of
lines with different directions and only few of them meet these criteria. This knowledge that only few of the transverse measurements are non-zero is
later used as a sparsity constraint in the electric field inverse problem.

\subsection{Discrete observation model}

\label{dfp} For the numerical evaluation, the domain is discretized and the electric field is represented as
\begin{equation}
\mathrm{e(x)} \approx \mathrm{e}_{N}(\mathrm{x}) = \sum_{i=1}^{N} \mathrm{e}_i \phi_i(\mathrm{x}),
\end{equation}
where  $\phi_i(\mathrm{x})$ are the chosen basis functions, $N$ is
the number of basis functions and $\mathrm{e}_i=(e_{ix},e_{iy})$
contains the electric field components. In the following, we denote
${e} = [e_{1x},e_{2x}, \cdots, e_{Nx}, e_{1y}, e_{2y}, \cdots,
e_{Ny}]^{\mathrm{T}}$ as the vector representation of
$\mathrm{e}_{N}(\mathrm{x})$.

The line integrals are evaluated along a set of straight lines which are formed by connecting pairs of points on the boundary $\partial \Omega$. In
practice, a finite number of points (or electrodes) are used for the measurements. If the number of electrodes is $n$, then the number of possible
line integral measurements is $m = n(n-1)/2$.

We stack the longitudinal line measurements into a vector $I^{\parallel} = [I^{\parallel}_1, \cdots, I^{\parallel}_m]^{\mathrm{T}} \in
\mathbb{R}^m$ and present the observations in a matrix form as
\begin{equation}\label{eq:SystemOfLinearEquations1}
I^{\parallel}={R}^{\parallel} {e}+\varepsilon,
\end{equation}
where ${R}^{\parallel}\in\mathbb{R}^{m\times 2N}$ is
called the longitudinal ray matrix and it consists of the integration coefficients of
the $m$ longitudinal line integrals, and $\varepsilon \in
\mathbb{R}^m$ is the measurement noise that is assumed to be random (Gaussian) white noise.

For the transverse line measurements, a similar matrix can be formulated. This matrix is called the transverse ray matrix $R^{\perp}\in\mathbb{R}^{m\times 2N}$ and it consists of the integration coefficients of the $m$ transverse line integrals. The numerical approximation of the ray matrices is described in detail in Section \ref{nm}.

\subsection{Discrete electric field inverse problem}

For the estimation of the field, the VT methods require
the values of both types of integrals for all possible lines. However, two
difficulties arise. The first one is that the transverse integral
measurements cannot be carried out using physical means. The second
one is that in practice we have only a finite number of lines and measurements available (limited data problem).
This means that there are areas in the
domain that are not covered by any of the line integrals, thus no information
can be retrieved from these areas.

Here, we deal with these problems by using two penalty terms.
We formulate the electric field inverse problem as a minimization
problem as follows
\begin{equation}\label{eq:MinimizationProblem}
{\hat{{e}}} = \min_{{e}} \{\|R^{\parallel}{{e}}-I^{\parallel}\|_2^2+\alpha\|R^{\perp}{{e}}\|_1+\beta\|W{{e}}\|_1\}.
\end{equation}
The first term is the data fidelity term, the second term is the $L_1$-norm of the transverse line integral measurements with a
\textcolor[rgb]{0.0,0.00,0.00}{regularization} parameter $\alpha
> 0$, and the third term is the $L_1$-norm of the discretized vector Laplace operator with a \textcolor[rgb]{0.00,0.00,0.00}{regularization} parameter $\beta > 0$. As discussed in Section
\ref{fs}, even though the transverse integrals cannot be measured directly, we can still say that only a small number of them are non-zero.
Therefore, we employ this knowledge by formulating an $L_1$-type sparsity prior that promotes such behaviour with the help of the transverse ray
matrix $R^{\perp}$.

To alleviate the limited data problem, we utilize the weighted vector Laplace operator. Loosely speaking, the vector Laplace operator
\cite{Grady2010} relates the local field values to the average of the surrounding points and thus imposes ``connectivity'' between the neighboring
points. The vector Laplace is also related to the current sources as
\begin{equation}
\nabla^2\mathrm{e} = \nabla(\nabla\cdot \mathrm{e})-
\nabla\times(\nabla\times \mathrm{e}) = \nabla(\nabla\cdot \mathrm{e})
= -\frac{1}{\sigma}\nabla(\nabla\cdot \mathrm{j}^s).
\end{equation}
Because of this and the sparsity of the current sources, we use also here the $L_1$-norm. Furthermore, because it is known that minimizing the
$L_1$-norm of the vector Laplace yields harmonic solutions that have their maxima on the boundary \cite{BurgerDepth}, we also use weighting factors
in (\ref{eq:MinimizationProblem}). \textcolor[rgb]{0.0,0.00,0}{ The discrete weighted laplace (in 2D) is defined as $W = \mathrm{w} (\Delta
\otimes I^{2\times 2})$. The weighted Laplace operator ensures connectivity between neighbouring nodes (local smoothness) and reduction of the depth
bias so that the maximum magnitude of the electric field will be correctly localized inside the domain \cite{Haufe2008,Pascual-Marqui,BurgerDepth}.
In our implementation, we employ the symmetric/normalized discrete Laplace operator $\Delta$ which is given by $\Delta=\mathrm{I}^{N\times
N}-\mathrm{diag}(H)^{-1/2} H\,\,\mathrm{diag}(H)^{-1/2}$, where $\mathrm{I}^{N\times N}$ is the identity matrix and the elements of matrix $H$ are}
\begin{equation}
\begin{array}{lcl}
 \mbox{if} \,\,\, i\neq j  \,\,\,  &H_{ij}& =\begin{cases}
- \frac{1}{d_{ij}} \, \mbox{, if $i$ and $j$ are connected with a vertex}
 \\0 \,\,\,\,\,\,\,\,\,\, \mbox{, otherwise}\end{cases}
\\  \mbox{if} \,\,\, i=j  \,\,\,  &H_{ii}& =  -\sum_j H_{ij}
\end{array}
\end{equation}
where $d_{ij}$ is the distance between nodes $i$ and $j$. The weights $w_i$ are the diagonal elements of the so-called resolution matrix
\cite{Pascual-Marqui} estimated in a similar way as in \cite{Haufe2008}. This resolution matrix is in our case given by $\Gamma=
K^{\mathrm{T}}(KK^{\mathrm{T}})^{-1}K$, where $K = (D^{\mathrm{T}}D)^{-1} D^{\mathrm{T}} R^{\parallel}$, and $D\in\mathbb{R}^{m\times n}$ is the
difference matrix for the potentials $u$ such as $I^{\parallel} = Du$. The matrix $\Gamma$ relates the minimum norm solution, $e_{\mathrm{MNE}}$,
with the actual field $e_{\mathrm{MNE}} = \Gamma e$ \cite{Pascual-Marqui}.

\subsubsection{Uniqueness of the solution}

For a unique reconstruction of an arbitrary vector field in a two dimensional domain, both the longitudinal and transverse measurements are required.
Numerically, this means that the null spaces of the longitudinal and transverse ray matrix do not coincide i.e.
$\mathcal{N}(R^\parallel)\cap\mathcal{N}(R^\perp)=\emptyset$ where $\mathcal{N}(\cdot)$ denotes the null space of a matrix.

Now let us assume that we are reconstructing an irrotational field that is a sum of two terms $\mathrm{e}(\mathrm{x}) = \mathrm{e}_b + \mathrm{e}_0$,
where $\mathrm{e}_b=-\nabla u_b$ is non-zero on the boundary and $e_0= - \nabla u_0$ has vanishing boundary values. This means that $R^\parallel e_0
= 0$ and $e_0\in\mathcal{N}(R^\parallel)$ where $e_0\in\mathbb{R}^{2N}$  is the vector representation of the field components. However, because
$\mathcal{N}(R^\parallel)\cap\mathcal{N}(R^\perp)=\emptyset$ we have that $\mathrm{e}_0\notin \mathcal{N}(R^\perp)$ unless the field is trivial i.e.
identically zero everywhere.

Now, by considering sparsity of the transverse integral, we implicitly impose that component $\mathrm{e}_0=0$ otherwise the equation $R^\perp
e\approx 0$ cannot hold. Thus, our formulation does not allow reconstruction of field components with vanishing values on the boundary. Therefore,
our solution can be considered as unique. In Section~\ref{sec:TransverseSimulation}, we show through simulation the
effect of the sparse transverse measurements on the solution.\\

\section{Numerical methods}

In this section, we describe how the numerical approximations of the line integrals and ray matrices were carried out and how the approach was tested
with numerical experiments.

\subsection{Numerical approximation of ray matrices}
\label{nm}

\begin{figure}[!htb]\centering
\includegraphics[width=14cm]{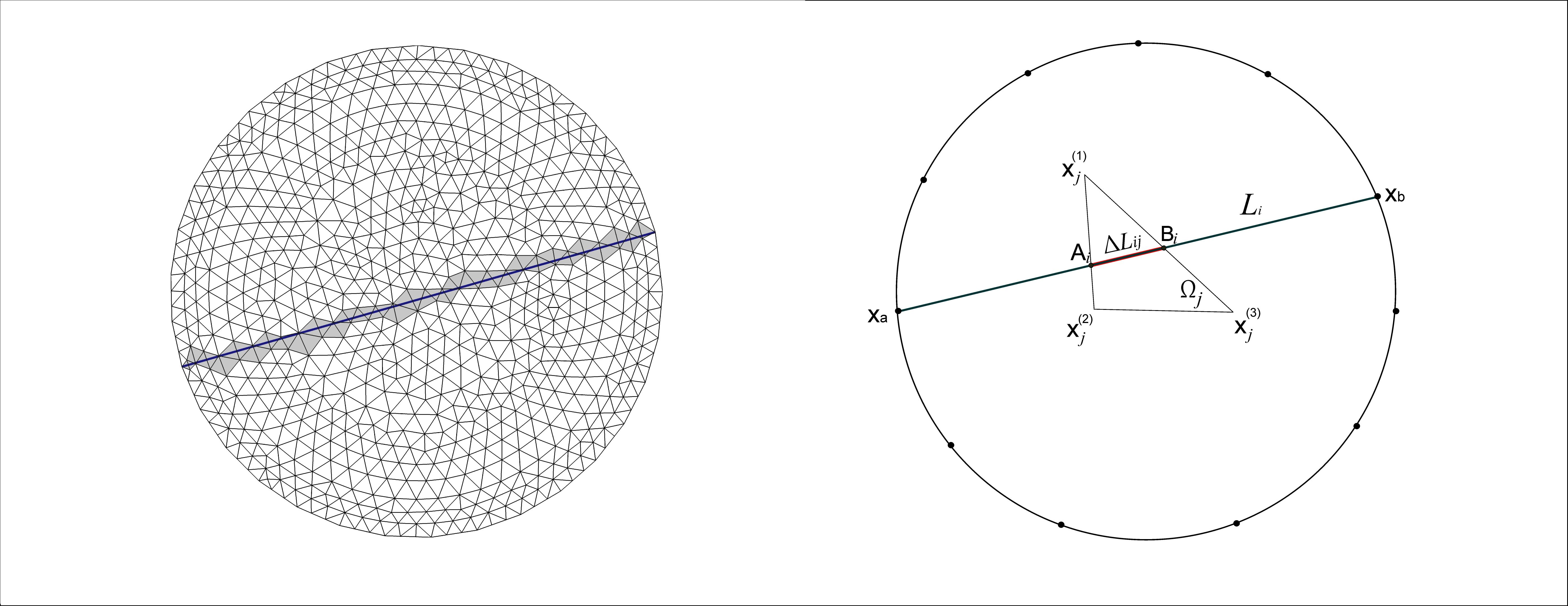}
\caption{Left: The computational domain is discretized into triangular elements, and the domain is traced with lines that connect pairs of points at the boundary. One of these lines $L_i$ passes through the elements that are colored with gray. Right: The line segment $\Delta L_{ij}$ is the intersection of the line $L_i$ and the triangular element $\Omega_j$. The coordinates of the points on this line segment between $A_i$ and $B_i$ can be expressed as a linear combination of the coordinates of the corner points of the triangle $\mathrm{x}_j^{(1)}$, $\mathrm{x}_j^{(2)}$ and $\mathrm{x}_j^{(3)}$.} \label{fig:illu}
\end{figure}

The domain $\Omega$ is divided into $N_E$ disjoint triangular elements, $\Omega=\cup_{j=1}^{N_E} \Omega_j$  with  $N$ nodes and the electric field is
expressed in a vector form as $e=[\mathrm{e}_x,\mathrm{e}_y]^\mathrm{T}\in \mathbb{R}^{2N}$ (as in Section \ref{dfp}). We use straight lines as the
integration paths. The same lines are used for both the longitudinal and transverse integrals. For the $i$th longitudinal measurement along the line
$L_i$, we can write
\begin{equation}
\label{eq:longitudlanLineIntegralapproximation} I^{\parallel}_i= \int_{L_i}\mathrm{e}(\mathrm{x})\cdot\mathrm{\hat{s}_i}~d\ell(\mathrm{x})=\sum_{j=1}^{N_E}
\int_{\Delta L_{ij}}\mathrm{e}(\mathrm{x})\cdot \mathrm{\hat{s}}_i~d\ell(\mathrm{x}),
\end{equation}
where $L_i = \sum_{j=1}^{N_E} \Delta L_{ij}$ gives the line segmentation and $\Delta L_{ij}=L_i\bigcap \Omega_j$ is the $j$th segment of $L_i$ (that is inside the element $\Omega_j$). Figure \ref{fig:illu} illustrates these variables. Note that $\Delta L_{ij}\neq \varnothing$ only if the line $L_i$ passes through the element $\Omega_j$.

\subsubsection{Line segment in an element}

Let us first examine a line segment $\Delta L_{ij}\neq\varnothing$ that passes through an element $\Omega_j$ that has corner points $\mathrm{x}_j^{(1)}$, $\mathrm{x}_j^{(2)}$ and $\mathrm{x}_j^{(3)}$ as shown in Figure \ref{fig:illu}.
A point on a line segment $\Delta L_{ij}$ has the
position vector $\mathrm{p}\in\mathbb{R}^2$
\begin{equation}\label{eq:lineParametric}
\mathrm{p}=\mathrm{x}^{A_i}_{j}+(\mathrm{x}_j^{B_i}-\mathrm{x}_j^{A_i})t = \mathrm{x}^{A_i}_{j}+\Delta\mathrm{x}_{ij}~t,
\end{equation}
where $\mathrm{x}^{A_i}_{j}$ and $\mathrm{x}^{B_i}_{j}$ are the
intersecting points between the line $L_i$ and the edges of element
$\Omega_j$ (see Figure \ref{fig:illu}),
$\Delta\mathrm{x}_{ij}=\mathrm{x}_j^{B_i}-\mathrm{x}_j^{A_i}$  and
$t~\in~[0,1]$. By changing variables, we obtain $\mathrm{\hat{s}}_i~d\ell(\mathrm{x})=
\Delta\mathrm{x}_{ij}dt$ and the line
integral in this element becomes
\begin{equation}\label{eq:longitudlanLineIntegralapproximation1}
I^{\parallel}_{ij} = \int_0^1
\Delta\mathrm{x}_{ij}\cdot\mathrm{e}(\mathrm{p}(t))~
dt=\|\Delta\mathrm{x}_{ij}\| \int_0^1
\mathrm{\hat{s}}_i\cdot\mathrm{e}(\mathrm{p}(t))~ dt,
 \end{equation}
as $\Delta\mathrm{x}_{ij}= \|\Delta\mathrm{x}_{ij}\| \mathrm{\hat{s}}_i$ where $\|.\|$ denotes the length of the vector.

The electric field values at the corner points of the element are
$\mathrm{e}^{(1)}_j$, $\mathrm{e}^{(2)}_j$ and $\mathrm{e}^{(3)}_j$.
We approximate the field inside the element using the linear
interpolation
\begin{equation}
\label{lif}
\mathrm{e(\mathrm{p}}(t))=\mathrm{e}^{(1)}_j+\begin{bmatrix}
\mathrm{e}^{(2)}_j-\mathrm{e}^{(1)}_j &
\mathrm{e}^{(3)}_j-\mathrm{e}^{(1)}_j\end{bmatrix}\mathrm{d}=\mathrm{e}^{(1)}_j+
K_j\mathrm{d},
\end{equation}
where $K_j = [ \mathrm{e}^{(2)}_j-\mathrm{e}^{(1)}_j ~~ \mathrm{e}^{(3)}_j-\mathrm{e}^{(1)}_j] \in \mathbb{R}^{2\times2}$ and $\mathrm{d} \in
\mathbb{R}^{2\times1}$ are the interpolation coefficients (or bary-centric coordinates) \cite{Coxeter69}. For the estimation of $\mathrm{d}$, we
employ an iso-parametric mapping in which the position vector in the element and the field are represented by the same interpolation polynomial
\cite{Vauhkonen1999}. In particular, a point $\mathrm{x}$ in $\Omega_j$ has position vector
\begin{equation}\label{eq:PointInElement}
\mathrm{x}=\mathrm{x}^{(1)}_{j}+\begin{bmatrix} \mathrm{x}^{(2)}_{j}-\mathrm{x}^{(1)}_{j} &
\mathrm{x}^{(3)}_j-\mathrm{x}^{(1)}_{j}\end{bmatrix}\mathrm{d}=\mathrm{x}^{(1)}_{j}+J_j\mathrm{d}
\end{equation}
where $\mathrm{d}=[d_{1}\,\ d_{2}]^\mathrm{T}$ with $d_1\geq0$, $d_2\geq 0$ and $d_1+d_2\leq 1$ and $J_j = [
\mathrm{x}^{(2)}_j-\mathrm{x}^{(1)}_j ~~
\mathrm{x}^{(3)}_j-\mathrm{x}^{(1)}_j] \in \mathbb{R}^{2\times2}$. Now $\mathrm{d}$ can by solved as
\begin{equation}
\label{dj}
\mathrm{d}={J}_j^{-1}(\mathrm{x}-\mathrm{x}^{(1)}_{j}).
\end{equation}
When we set $\mathrm{x}=\mathrm{p}$ and combine Equations (\ref{eq:lineParametric}) and (\ref{dj}), we can estimate the interpolation coefficient on the line segment with respect to $t$
\begin{equation}
\mathrm{d}={J}_j^{-1}(\mathrm{x}^{A_i}_{j}+\Delta\mathrm{x}_{ij}~t-\mathrm{x}^{(1)}_{j}).
\end{equation}
Now, this results in
\begin{equation}
\mathrm{e(\mathrm{p}}(t))=\mathrm{e}^{(1)}_j+
K_j{J}_j^{-1}(\mathrm{x}^{A_i}_{j}+\Delta\mathrm{x}_{ij}~t-\mathrm{x}^{(1)}_{j}),
\end{equation}
and Equation (\ref{eq:longitudlanLineIntegralapproximation1}) becomes
\begin{equation}\label{eq:longitudlanLineIntegralapproximatio4}
\begin{split}
I^{\parallel}_{ij}=&
\|\Delta\mathrm{x}_{ij}\|\left(\mathrm{\hat{s}}_i\cdot\mathrm{e}_j^{(1)}+\int_0^1
\mathrm{\hat{s}}_i\cdot
K_j{J}_j^{-1}(\mathrm{x}^{A_i}_{j}+\Delta\mathrm{x}_{ij}~t-\mathrm{x}^{(1)}_{j})~ dt\right) \\
=& \|\Delta\mathrm{x}_{ij}\|~\mathrm{\hat{s}}_i
\cdot\left(\mathrm{e}_j^{(1)}+K_jJ_j^{-1}(\frac{1}{2}\Delta\mathrm{x}_{ij}-\mathrm{x}^{(1)}_j)
\right).
\\
\end{split}
\end{equation}
Finally when we write $K_j$ explicitly and denote $C_{ij}=J_j^{-1}(\frac{1}{2}\Delta\mathrm{x}_{ij}-\mathrm{x}^{(1)}_j)=[c_{1},c_{2}]^{\mathrm{T}}
\in \mathbb{R}^{2\times1}$ we get
\begin{equation}\label{eq:fin}
I^{\parallel}_{ij}=\|\Delta\mathrm{x}_{ij}\|~\mathrm{\hat{s}}_i
\cdot \left[ (1-c_1-c_2)\mathrm{e}_j^{(1)}+c_1\mathrm{e}_j^{(2)}+c_2\mathrm{e}_j^{(3)}
\right].
\end{equation}
We use Equation (\ref{eq:fin}) to create a procedure for constructing the ray matrices.

\subsubsection{Construction of ray matrices}

The ray matrices are used to operate on the electric field in order to obtain the line integral measurements. We denote the relationship between the
longitudinal measurements and the ray matrix operator as follows
\begin{equation}
I^{\parallel} =R^{\parallel}e=\left[ R_x^{\parallel} ~~ R_y^{\parallel} \right] \left[ \begin{array}{c} {e}_x \\ {e}_y \end{array} \right].
\end{equation}
Data $I^{\parallel}=\left[ I_1, \cdots, I_m \right]^{\mathrm{T}} \in \mathbb{R}^m$ contains all the longitudinal line measurements where $m$ is the
number of the line integrals that equals to $m=\frac{n(n-1)}{2}$ where $n$ is the number of measurement electrodes on the boundary. Matrix
$R^{\parallel} \in \mathbb{R}^{m\times2N}$ is the longitudinal ray matrix operator and $N$ is the number of nodes of the discretized domain.
$R^{\parallel}$ consists of $R_x^{\parallel} \in \mathbb{R}^{m\times N}$ and $R_y^{\parallel}\in\mathbb{R}^{m\times N}$  that contain the
contributions of the integral coefficients of the $x$ and $y$ field components, respectively. The procedure that was used to construct the
longitudinal ray matrix is shown in Table \ref{rmc}. A similar procedure can be carried out to construct the transverse ray matrix $R^{\perp}$ by
exchanging the vector $\mathrm{\hat{s}}_i$ with $\mathrm{\hat{s}}_i^\perp$ in the first step.


\begin{table}[hbtp]
\caption{Procedure to construct the longitudinal ray matrix.}
\begin{center}
\begin{small}
\begin{tabular}{l}
\hline
\textbf{for} {\it i}=1:{\it m}, go through all the integration lines\\

\qquad take line $L_i$ and determine the corresponding unit vector $\hat{s}_i=(s_x,s_y)$ along the line.\\

\qquad\qquad \textbf{for} j=1:$N_E$, go through all the elements\\

\qquad\qquad\qquad take element $\Omega_j$ that has corner points $\mathrm{x}_j^{(1)}$, $\mathrm{x}_j^{(2)}$ and $\mathrm{x}_j^{(3)}$, where \\
\qquad\qquad\qquad $j^{(1)}$, $j^{(2)}$ and $j^{(3)}$ are the corresponding node indices.\\

\qquad\qquad\qquad\qquad \textbf{if $\Delta L_{ij}=L_i \cap \Omega_j=\varnothing$}\\
\qquad\qquad\qquad\qquad\qquad go to the next element.\\

\qquad\qquad\qquad\qquad\textbf{else}   \\
\qquad\qquad\qquad\qquad\qquad calculate $\Delta \mathrm{x}_{ij}$ and $\| \Delta \mathrm{x}_{ij} \|$. \\
\qquad\qquad\qquad\qquad\qquad calculate $J_j$ and $C_{ij}=[c_1,c_2]^{\mathrm{T}}$.\\
\qquad\qquad\qquad\qquad\qquad update the following entries of the $i$th row of the ray matrix.\\
\\
\qquad\qquad\qquad\qquad\qquad  $R^{\parallel}(i,j^{(1)})=R^{\parallel}(i,j^{(1)})+\|\Delta \mathrm{x}_{ij}\| s_x (1-c_1-c_2)$\\
\qquad\qquad\qquad\qquad\qquad  $R^{\parallel}(i,j^{(2)})=R^{\parallel}(i,j^{(2)})+\|\Delta \mathrm{x}_{ij}\| s_x c_1$\\
\qquad\qquad\qquad\qquad\qquad  $R^{\parallel}(i,j^{(3)})=R^{\parallel}(i,j^{(3)})+\|\Delta \mathrm{x}_{ij}\| s_x c_2$\\
\qquad\qquad\qquad\qquad\qquad  $R^{\parallel}(i,j^{(1)}+N)=R^{\parallel}(i,j^{(1)}+N)+\|\Delta \mathrm{x}_{ij}\| s_y (1-c_1-c_2)$\\
\qquad\qquad\qquad\qquad\qquad  $R^{\parallel}(i,j^{(2)}+N)=R^{\parallel}(i,j^{(2)}+N)+\|\Delta \mathrm{x}_{ij}\| s_y c_1$\\
\qquad\qquad\qquad\qquad\qquad  $R^{\parallel}(i,j^{(3)}+N)=R^{\parallel}(i,j^{(3)}+N)+\|\Delta \mathrm{x}_{ij}\| s_y c_2$\\

\\
\qquad\qquad\qquad\qquad \textbf{end}\\

\qquad\qquad\textbf{end} \\
\textbf{end} \\

\hline

\end{tabular}
\end{small}
\end{center}
\label{rmc}
\end{table}


\subsection{Numerical experiments}

\textcolor[rgb]{0,0,0}{In the experiments, we study electric fields generated by dipole sources in a bounded 2-dimensional circular domain with homogeneous electrical conductivity $\sigma=1$ S/m.} We note that the same methods can also be applied to any other convex domain. The domain contained $n=32$
equally spaced electrodes around the boundary. Lines for the integration were formed by connecting all pairs of electrodes which resulted in total of
$m=496$ lines. Two computational meshes were used: the finer one consisted of $\bar{N}_E=9721$ triangular elements linking $\bar{N}=3045$ nodes, and
the coarser one of $N_E=1408$ triangular elements with $N=760$ nodes.

\subsubsection{Simulated forward fields and integral data}

For the estimation of the longitudinal integral data and the forward electric field, the finer mesh was used. First, a single dipole source was
selected and the corresponding scalar electric potential distribution $\bar{u}\in\mathbb{R}^{\bar{N}}$ was computed by solving the Poisson problem
(\ref{p1}) with boundary conditions (\ref{p2}) and (\ref{p3}) using finite element method (FEM) with linear nodal basis functions \cite{Wolters2004}.
The dipole source function was numerically approximated using the mathematical dipole model \cite{Schimpf2002}. The longitudinal integral
observations were calculated by taking the differences of the potential values at the electrodes that locate at the ends of the integration lines.
Gaussian white noise was added to the data using two signal-to-noise ratios, 40 dB and 20 dB. The signal-to-noise ratio (SNR) is given by
\begin{equation}
\mathrm{SNR}=20\,\mathrm{log}_{10}\frac{\|I^{\parallel}\|_2}{\|\varepsilon\|_2}
\end{equation}

The forward field, given by $\mathrm{\bar{e}}=-\nabla \bar{u} $, was estimated numerically by applying the linear gradient reconstruction approach
\cite{Correa2011}. The forward field ${\bar{{e}}}\in\mathbb{R}^{2\bar{N}}$ was then projected to the inverse mesh ${e}=P\bar{{e}}$, where $P\in
\mathbb{R}^{2N\times 2\bar{N}}$ is a linear reduction mapping operator, in order to be able to compare the reconstructed field with the correct one.

\subsubsection{Electric field reconstructions}

The electric fields were reconstructed using the coarse mesh. First, the longitudinal and transverse ray matrices were constructed for the mesh by
using the procedure described in Section \ref{nm}. Then, the non-linear minimization problem (\ref{eq:MinimizationProblem}) was solved using convex
optimization techniques and more precisely CVX toolbox (SDP3 solver) \cite{Boyd2004,cvx,gb08}. The regularization parameters were kept constant in
all experiments ($\alpha=0.06$ and $\beta=0.016$ in our case). The $\beta$ value was used to scale the coefficients of the discrete weighted Laplace
operator (that promotes connectivity between neighbouring nodes) to match the average distance between the mesh nodes. The choice of $\alpha$ was
carried out empirically, and it was found that $\alpha$ value has to be 3--4 times higher than $\beta$ to ensure that the orientations of the field
lines are estimated properly.

\subsubsection{Reconstruction error metrics}

For the evaluation of the reconstructions, we used two different measures. First, the average magnitude ratio (MR) between the reconstructed and the actual field was computed as follows
\begin{equation}
\mathrm{MR} = \frac{1}{N} \sum_{i=1}^{N} \frac{\sqrt{\hat{e}_{ix}^{2}+\hat{e}_{iy}^{2}}}{\sqrt{{e}_{ix}^{2}+{e}_{iy}^{2}}}.
\end{equation}
\noindent
The closer the MR is to one the better the fields match with respect to the magnitude.

Second, the average cosine similarity (CS) was used to quantify the difference between the directions of the reconstructed and the actual field
\begin{equation}
\mathrm{CS} = \frac{1}{N} \sum_{i=1}^{N} \mathrm{cos}(\mathrm{\hat{e}}_i,\mathrm{e}_i)= \frac{1}{N} \sum_{i=1}^{N} \frac{\hat{e}_{ix}e_{ix}+\hat{e}_{iy}e_{iy}}{\sqrt{\hat{e}_{ix}^{2}+\hat{e}_{iy}^{2}}\sqrt{{e}_{ix}^{2}+{e}_{iy}^{2}}}.
\end{equation}
\noindent CS has a value between --1 and +1, and the closer the value is to one the better the directions of the fields match. Moreover, CS is close
to zero when the reconstructed and the actual field are perpendicular to each other, and finally CS is close to --1 when the directions of the fields
are opposite.


\section{Results and discussion}

\subsection{Effect of $L_1$-norm constraint of transverse measurements}\label{sec:TransverseSimulation}

\begin{figure}[!htb]\centering
\includegraphics[width=14cm]{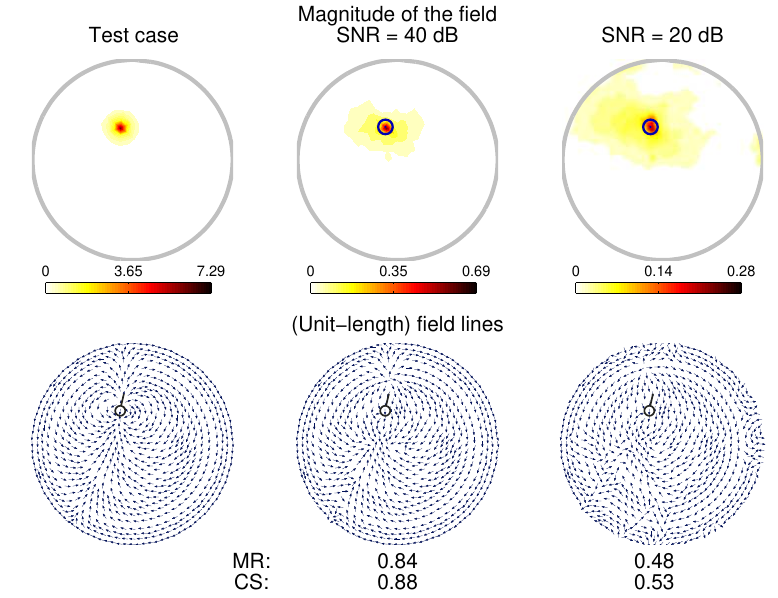}
\caption{\textcolor[rgb]{0,0,0}{The left column shows a test case from which the boundary data was extracted. The middle column shows the
reconstructed magnitude distribution and normalized electric field lines when the $L_1$ sparsity of transverse measurements was used ($\alpha=0.06$).
The right column shows the corresponding reconstruction when this constraint was not used ($\alpha=0$). It can be seen that by omitting the $L_1$
sparsity the quality of the reconstruction is reduced, especially the directions of the electric field lines exhibit significant
errors.}}\label{comppic}
\end{figure}

Figure \ref{comppic} illustrates the benefit of using the $L_1$ constraint of the transverse measurements. The first column shows the true magnitude
of the electric field that is generated by a dipole source and the corresponding unit-length field lines with the location and orientation of the
underlying dipole source (circle). The last two columns show the reconstructed electric field magnitude and field lines from noiseless boundary data
with and without the $L_1$ constraint, respectively.

As we can see the magnitude distributions of the reconstructed fields have similar patterns as the true field; however, the orientations of the field
lines show significant errors when the $L_1$ constraint is omitted. The same can be seen from the magnitude ratio numbers which are quite similar,
0.88 with and 0.75 without the constraint, and the cosine similarity numbers which decrease drastically from 0.90 to 0.31 when the $L_1-$norm is not
used.

The dipole can be viewed as a positive and an equivalent negative charge that are separated by a vanishingly small distance. The corresponding field
lines point outwards from the positive and inwards from the negative end of the dipole. From the reconstructed field lines with the $L_1$ constraint,
it can be seen that the locations of the positive and the negative charge can be found but they are separated by a small non-zero distance. When the
$L_1$ constraint is not considered these locations cannot be found at all.

The transverse measurements can be interpreted as fluxes across the integration lines. For an electric field generated by a dipole source, the total
flux across most of the integration lines is zero as already discussed in Section~\ref{fs} and Figure~\ref{fig:LineOrthogonalToVectorMoment}. The
sparsity constraint of transverse measurements exactly ensures this. Consequently, it first forces the field lines to orientate similar to the field
of closely separated positive and negative charge. Second, it forces that the flux across line integrals close to the boundary is zero (see for
example dash line of integration in Figure~\ref{fig:LineOrthogonalToVectorMoment}) thus making the field tangential on the boundary. From the
mathematical point of view, based on the analysis in Section~\ref{sec:TransverseThroughRadon}, the sparsity constraint for the transverse
measurements implies focal activity and homogeneous Neumann conditions. Of course these effects are limited by the discretization level of the
domain, the numerical approximations of the line integrals, the number of measurements and the measurement noise as we shall see in the next Section.
\\

\subsection{Reconstructions in the presence of noise}

\begin{figure}[!htb]\centering
\includegraphics[width=14cm]{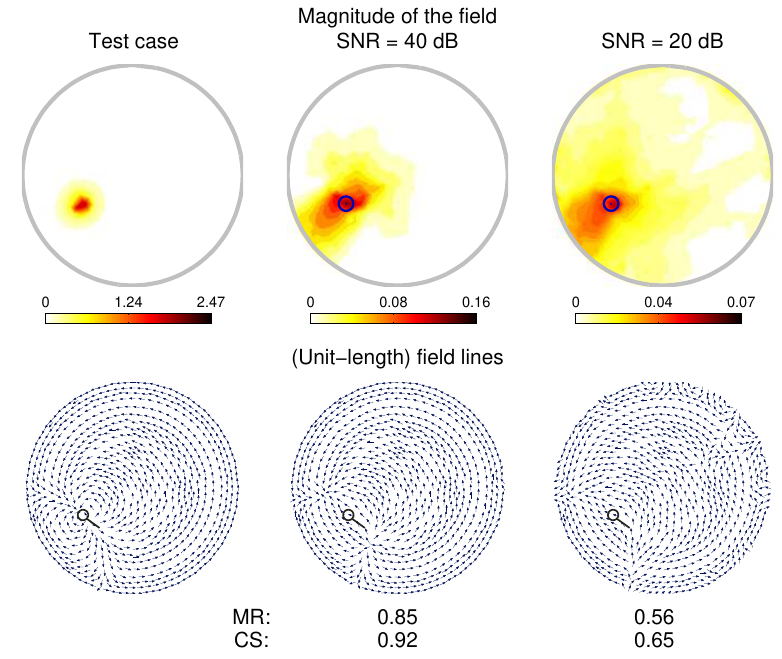}
\caption{First column: Top picture shows the magnitude of the electric field that is generated by a radial dipole source.
Bottom picture shows the corresponding unit length field lines and the location and orientation of the underlying dipole source. Second column:
Top picture shows the average electric field magnitude that is calculated over 10 reconstructions and the circle shows the correct location of the underlying dipole source. The 10 reconstructions were computed using 10 different realizations of noisy boundary data with SNR\,=\,40\,dB. Bottom picture shows the corresponding unit-length electric field lines. Third column: Top and bottom pictures show similarly the reconstructed electric field when SNR\,=\,20\,dB. The average magnitude ratio (MR) and cosine similarity (CS) values are given under the reconstructions.} \label{fig:sup1}
\end{figure}

\begin{figure}[!htb]\centering
\includegraphics[width=14cm]{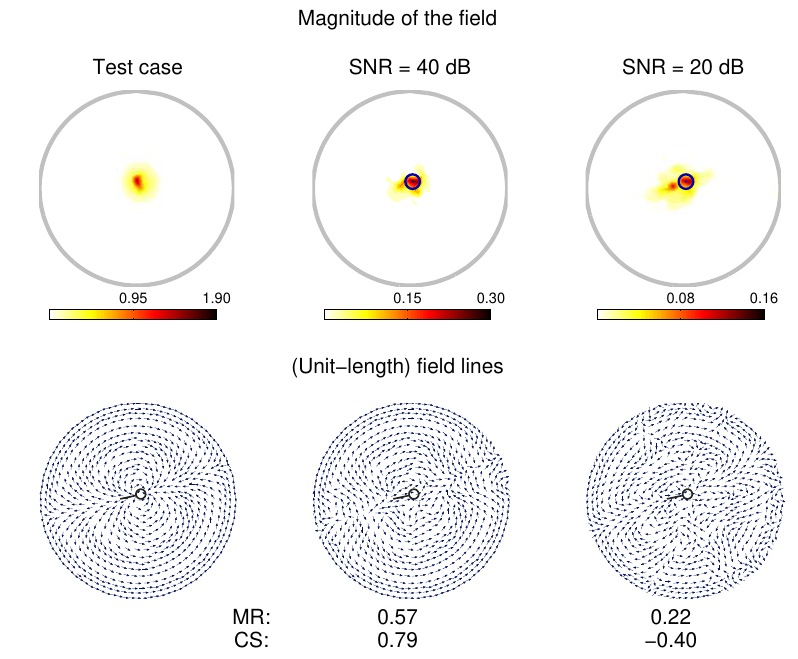}
\caption{First column: Top picture shows the magnitude of the electric field that is generated by a tangential dipole source. Bottom picture shows the corresponding  unit length electric field lines with the underlying dipole source. The other columns show the electric field reconstructions as explained in Figure \ref{fig:sup1}.}
\label{fig:sup2}
\end{figure}

\begin{figure}[!htb]\centering
\includegraphics[width=14cm]{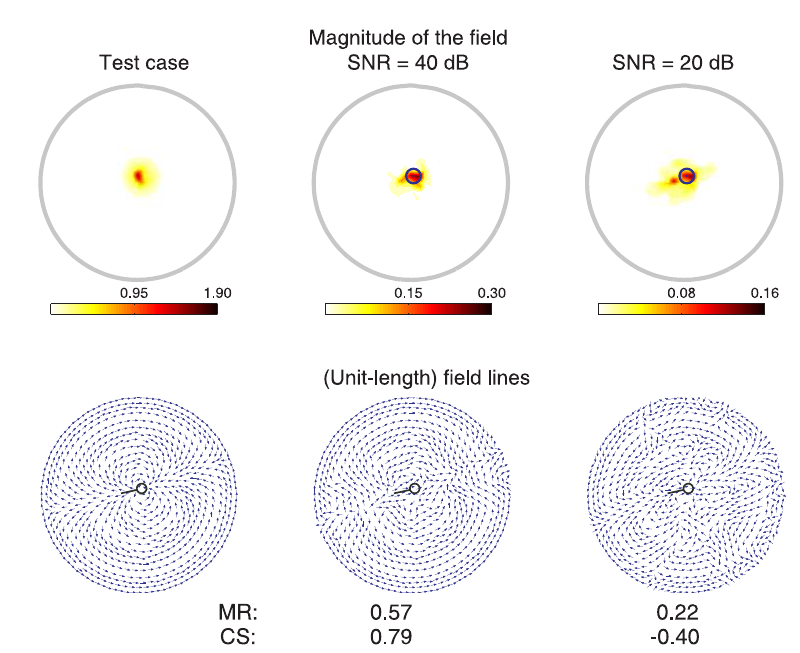}
\caption{First column: Top picture shows the magnitude of the electric field that is generated by a dipole source in the centre of the domain. Bottom picture shows the corresponding  unit length electric field lines with the underlying dipole source. The other columns show the electric field reconstructions as explained in Figure \ref{fig:sup1}.}
\label{fig:cen1}
\end{figure}

Figures \ref{fig:sup1}--\ref{fig:cen1} show the reconstruction results of the electric fields produced by a single dipole source which has radial
orientation (Figure~\ref{fig:sup1}), tangential orientation (Figure~\ref{fig:sup2}) and is located in the centre of the domain
(Figure~\ref{fig:cen1}). In the following Figures, the last two columns show the average electric field magnitude and field lines that are calculated
over 10 reconstructions. The 10 reconstructions were computed using 10 different realizations of noisy boundary data with SNR\,=\,40\,dB (second
column) and SNR\,=\,20\,dB (third column).

From the upper row of the figures (showing the field magnitude), we see that the locations where the reconstructed fields get their maximum value are
very close to the correct locations (of the source) in all the test cases: in fact, the location is exactly the same in the test cases shown in
Figures \ref{fig:sup1} and \ref{fig:sup2}, and the location differences in Figure \ref{fig:cen1} correspond to the distance of a single node in the
mesh.

From the reconstructed field lines in the lower row of Figures \ref{fig:sup1} and \ref{fig:sup2}, it can be seen that the locations of the positive
and the negative charge are found correctly but that they are separated by a small non-zero distance. When the dipole source is in the centre of the
domain, Figure \ref{fig:cen1}, the locations of the positive and negative charge are not evident from the field lines.

The maximum values of the reconstructed fields are lower by an order of magnitude when compared to the actual ones. However, in the low noise cases (SNR = 40 dB) in Figures \ref{fig:sup1} and \ref{fig:sup2}, the magnitude ratio values are still very high which indicates that the magnitude errors are present only near the dipole sources and elsewhere the magnitudes are reconstructed accordingly. Similarly in the field orientations, there are differences merely close to the dipole sources, and especially for the low noise cases the CS values are high which indicates that the field orientations are correct in most parts of the domain. These reconstruction errors close to the source were expected because the dipole source causes a discontinuity in the field.

By comparing the test cases, we can observe that the magnitude and orientation errors are larger when the dipole is deep in the domain than close to
the boundary. We also see that the reconstruction accuracy decreases with increasing noise as expected.
\subsection{Multiple sources}
As we saw, the proposed approach gives stable estimates for both the magnitude and the orientation of the electric field when it is generated by a
single focal source. It can be said that the same L1-penalty terms are also valid for multiple source cases, however, it seems that further
information on the structure of the field is required for stable reconstructions.
\begin{figure}[!htb]
\centering
\includegraphics[width=9cm]{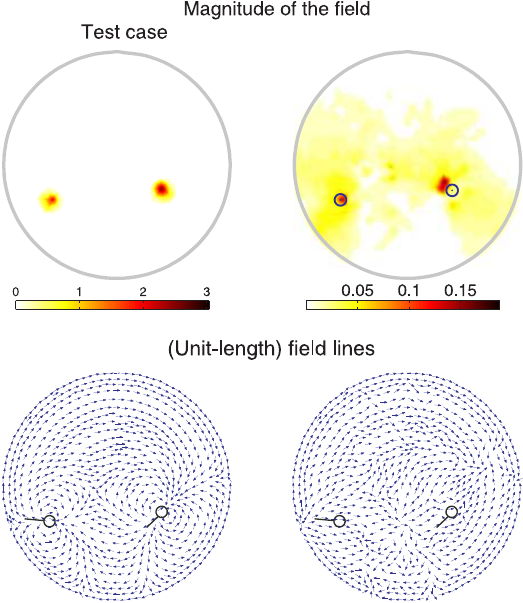}
\caption{First column: Top picture shows the magnitude of the electric field that is generated by two dipole sources. Bottom picture shows the
corresponding unit length electric field lines with the underlying dipole sources. The right column shows the electric field reconstruction.}
\label{fig:2sourceCase}
\end{figure}
As an example, we show in Figure~\ref{fig:2sourceCase} a preliminary result of a two source case. As can be seen, the magnitude can still be
recovered surprisingly well considering the limited amount of (measurement and prior) information available: for example, the locations of the
sources can be determined based on the highest magnitude values. The field orientations, on the other hand, show more errors here than in the one
source cases.

\FloatBarrier

\section{Conclusions and future work}

This was a proof-of-concept study to characterize non-zero divergence vector fields in a bounded domain using the longitudinal measurements and
appropriately chosen L1-penalty terms. To the best of our knowledge, this is the first time that numerical reconstructions have been presented under
these conditions. This type of problem is of great interest due to its many applications, especially in the EEG source imaging. As a comparison to
the widely used source imaging methods, we argue that the VT framework could be beneficial because it does not require an explicit mathematical model
for the sources.

We first showed that the longitudinal measurements are directly determined by the electric potentials at the boundary and the transverse measurements
are related to the underlying current sources. We explained that even though the transverse measurements cannot normally be physically measured, they
still can be utilized in the electric field inverse problem as a penalty term.

Our numerical test cases included reconstructions of non-zero divergence electric fields generated by a focal source with varying direction and
location. We showed that the pattern of the electric field magnitude could be reconstructed correctly using VT, even though there were errors near
the dipole source. Nevertheless, for example, the correct location of the source activity could be determined based on the reconstructed field
magnitudes. Also, the reconstructed field lines follow similar trajectories to the real ones with some deviations only near the dipole sources. These
reconstruction errors were expected because the field is discontinuous at the dipole source.

Therefore, we conclude that our approach is able to give stable estimates for single focal source cases, however, the recovery of a field generated
by multiple sources requires further research. Nevertheless, we also note that our preliminary result for a two-source case was promising. In
addition to multiple source cases, in the future we shall extend the proposed approach for 3-dimensional VT problems with non-homogeneous material
properties.

\section*{Acknowledgment}

The authors would like to thank the John S. Latsis Public Benefit Foundation and the Finnish Cultural Foundation for financial support.

\section*{References}
\bibliography{vft_bibliography_02_v1}


\section*{Appendix: Electric field decomposition and line integrals}

According to Helmholtz decomposition \cite{Arfken2005}, any vector field can be decomposed into a sum of irrotational (curl-free) and solenoidal (divergence-free) component. When the field is zero on the domain boundary,
the irrotational and solenoidal components are unique, and they can be recovered from the
transverse and longitudinal measurements, respectively \cite{Braun1991,Prince1994}. However, when non-zero boundary conditions are present the decomposition is not unique anymore. For this case, however, a unique decomposition can be found by adding a harmonic
component which is both irrotational and solenoidal into the sum \cite{Braun1991,Osman1997}.

Let's consider a similar electric field $\mathrm{e: \Omega \rightarrow \mathbb{R}^2}$ in a bounded domain as in Section \ref{th1}. We set
$\mathrm{e}|_{\partial \Omega} \neq 0$ and $\mathrm{e} \cdot \hat{n} |_{\partial \Omega}=0$ on the boundary. The non-zero boundary conditions imply
that the decomposition of the field is of the form
\begin{equation}
\mathrm{e} = \mathrm{e}_\mathrm{I}+\mathrm{e}_\mathrm{S}+\mathrm{e}_\mathrm{H},
\end{equation}
where $\mathrm{e}_\mathrm{I}$, $\mathrm{e}_\mathrm{S}$ and $\mathrm{e}_\mathrm{H}$ are the irrotational, solenoidal and harmonic component, respectively. The following properties apply for the irrotational and solenoidal components $\nabla \cdot \mathrm{e} = \nabla \cdot \mathrm{e}_\mathrm{I}$, $\nabla \times \mathrm{e} = \nabla \times \mathrm{e}_\mathrm{S}$ and the harmonic component satisfies both $\nabla \cdot \mathrm{e}_\mathrm{H} = 0$, $\nabla \times \mathrm{e}_\mathrm{H} = 0$ and in addition also the boundary conditions.

Under the quasi-static approximation, the electric field is irrotational. It follows that the irrotational component can be expressed using a scalar
potential $\mathrm{e}_\mathrm{I}=-\nabla q$ and the solenoidal component vanishes $\mathrm{e}_\mathrm{S}=0$. Furthermore, we write the harmonic
component as $\mathrm{e}_\mathrm{H}=-\nabla r + \nabla \times q$ where $\nabla \times = (\frac{\partial}{\partial y},-\frac{\partial}{\partial x})$
corresponds to the 2-dimensional curl-operator. Now, $\nabla r = 0$ due to the boundary conditions. Therefore, the decomposition gets the form
\begin{equation}
\mathrm{e}(\mathrm{x})=-\nabla
q(\mathrm{x})+\nabla \times p(\mathrm{x}).
\end{equation}

The Radon transform of $\mathrm{e(x)}$ can be expressed using the Radon property (\ref{eq:Radon_derivative}) as
\begin{equation}
\begin{split}
\tilde{\mathrm{e}}_{\Omega}=\mathcal{R}_{\Omega}\{\mathrm{e}\}(l,\mathrm{s}^{\perp}) &= - \mathrm{\hat{s}}^{\perp}~\frac{\partial}{\partial
l}\mathcal{R}_{\Omega}(q)+\mathrm{\hat{s}}~ \frac{\partial}{\partial l}\mathcal{R}_{\Omega}(p),
\end{split}
\end{equation}
where ${\hat{s}}^{\perp}$ is the unit normal vector perpendicular and $\hat{s}$ along the line $L$. Now, the longitudinal
(\ref{eq:lineIntegralPreviousWork}) and transverse (\ref{eq:lineIntegralPreviousWork_transverse}) integral have the form
\begin{eqnarray}
I^\parallel &=& \mathrm{\hat{s}}\cdot \mathcal{R}_{\Omega}\{\mathrm{e}\}
(l,\mathrm{s}^{\perp})= \frac{\partial}{\partial l}\mathcal{R}_{\Omega}(p)\label{eq:LongitudinalMeasuremnt_Pot} \\
I^{\perp}&=& \mathrm{\hat{s}}^\perp\cdot
\mathcal{R}_{\Omega}\{\mathrm{e}\}(l,\mathrm{s}^{\perp})=
\frac{\partial}{\partial
l}\mathcal{R}_{\Omega}(q).\label{eq:TranverseMeasurement_Pot}
\end{eqnarray}

From these measurements, the electric field components can be solved as
\begin{eqnarray}
p &=&\mathcal{R}^{-1}\{\tilde{p}\} =
\frac{1}{4\pi}\mathcal{R}^{\#}\mathcal{H}\frac{\partial}{\partial
l}\mathcal{R}\{p\} =\frac{1}{4\pi}\mathcal{R}^{\#}\mathcal{H}
I^{\parallel}
\\
q &=&\mathcal{R}^{-1}\{\tilde{q}\} =
\frac{1}{4\pi}\mathcal{R}^{\#}\mathcal{H}\frac{\partial}{\partial
l}\mathcal{R}\{q\} =\frac{1}{4\pi}\mathcal{R}^{\#}\mathcal{H}
I^{\perp}.
\end{eqnarray}
Thus, both types of line integral measurements are needed for the full recovery of an
electric field in a bounded domain with non-zero boundary conditions. $I^{\parallel}$ is associated solely with the harmonic component, thus the boundary conditions, and $I^{\perp}$ is needed to recover the irrotational component.

\end{document}